\documentclass[runningheads, ]{llncs}

\usepackage{graphicx}
\usepackage{sgame}
\usepackage{booktabs}
\usepackage{amsfonts}
\usepackage{amssymb}
\usepackage{todonotes}
\usepackage{algorithm}
\usepackage[noend]{algpseudocode}
\usepackage{mathtools}
\usepackage{float}

\newcommand{\ubar}[1]{\text{\b{$#1$}}}

\algnewcommand{\Initialize}[1]{%
  \State \textbf{Initialize:}
  \Statex \hspace*{\algorithmicindent}\parbox[t]{.8\linewidth}{\raggedright #1}
}

\begin{document}
\title{The Price of Pessimism for Automated Defense
\thanks{Funded by the Auerbach Berger Chair in Cybersecurity held by Spiros Mancoridis, at Drexel University}}
\titlerunning{The Price of Pessimism}
\author{Erick Galinkin\inst{1}\orcidID{0000-0003-1268-9258} \and
Emmanouil Pountourakis\inst{1}\orcidID{0000-0002-5023-1099} \and
Spiros Mancoridis\inst{1}\orcidID{0000-0001-6354-4281}}
\authorrunning{E. Galinkin et al.}
\institute{Drexel University, Philadelphia PA 19104, USA \\
\email{eg657@drexel.edu}
}

\maketitle

\begin{abstract}
The well-worn George Box aphorism ``all models are wrong, but some are useful'' is particularly salient in the cybersecurity domain, where the assumptions built into a model can have substantial financial or even national security impacts.
Computer scientists are often asked to optimize for worst-case outcomes, and since security is largely focused on risk mitigation, preparing for the worst-case scenario appears rational.
In this work, we demonstrate that preparing for the worst case rather than the most probable case may yield suboptimal outcomes for learning agents. 
Through the lens of stochastic Bayesian games, we first explore different attacker knowledge modeling assumptions that impact the usefulness of models to cybersecurity practitioners.
By considering different models of attacker knowledge about the state of the game and a defender's hidden information, we find that there is a cost to the defender for optimizing against the worst case.
\end{abstract}
\section{Introduction}
Cybersecurity incidents continue to grow in frequency and volume each year, and are estimated to cause \$8 billion worth of damage in 2023~\cite{brooks2023cybersecurity}.
While cybersecurity awareness continues to grow and companies continue to invest in their security functions, the majority of threat response functions are still carried out manually by cybersecurity analysts.
As a result, there is a move to automate parts of cyber threat response -- something clearly illustrated by the wide availability and marketing of security orchestration, automation, and response (SOAR) systems. 
These systems, however, are largely rule-based -- they take some specific action when some specific criterion is met.
This is because SOAR lacks any direct knowledge of the network system it is responsible for helping defend.
We instead consider the use of attacker simulation to train defensive agents and aim to answer the question of what assumptions about attacker information should be made to train the most robust defensive agents.

Abstractions of complex systems generally trade accuracy for tractability because there is some use in modeling that system in different scenarios. 
In cybersecurity, we use modeling to consider potential future states of the system we are tasked with defending and how different threats and risks~\cite{Shostack2014ThreatModeling} may be realized.
As an additional constraint in cybersecurity modeling, accuracy and timeliness are constantly in tension. 
Responding inaccurately to a potentially malicious event that was triggered by benign behavior can have a significant cost, but not responding fast enough to a malicious event can be even more expensive.
In the automation of these responses, there is a tendency to optimize for the worst-case.
However, as we know from \textit{e.g.} linear programming~\cite{illes2002pivot}, there are cases where a solution that is worst-case optimal is empirically suboptimal in practice.
Our work seeks to understand the ``price of pessimism'' -- that is, the cost to a computer network defender for overestimating the knowledge or capability of an attacker.

While a number of different game types have been used across the security games space~\cite{sokri2020game,Liang2013GameSecurity}, a small handful comprise the preeminent models used in the space -- The Bayesian leader-follower game\footnote{Also known as the Stackelberg game}, the stochastic Bayesian game, and the two player zero sum game. 
When training a learning agent for cyberdefense, decisions about attacker modeling directly impact the policy learned by the defending agent. 
We concern ourselves primarily with the two Bayesian game variants under differing prior knowledge and observability assumptions where the attacker is an actor who seeks to deploy ransomware across a target network.
While the impact of altering a game's state or action space is clear and the implications are well-understood, there are other modeling assumptions that are often made implicitly.
These assumptions are critical parts of the game design that impact its usefulness to cybersecurity practitioners.
In this work, we make assumptions explicit about the presence of attackers and noise, and consider the impact of assumptions about attacker knowledge.

Our work begins with a presentation of related work in the field of security game theory.
We then define the stochastic Bayesian game model and setting we use as an abstraction of attacker-defender interaction on a computer network before presenting a theoretical model of how defender belief about attacker knowledge is liable to influence their behavior.
In this manuscript, we are particularly interested in the case we deem most realistic -- one where the attacker has limited knowledge of the target network.
Our model, being built on partially observable stochastic Bayesian games which have no known general solution concept~\cite{galinkin2023simulation}, requires that we develop a decision theory for the players. 
To overcome the limitations of an attacker to optimize against a particular defender, we introduce the use of the restricted Bayes/Hurwicz criterion~\cite{ellsberg2015risk} for decision making under uncertainty.
In order to validate our theoretical findings, we leverage reinforcement learning in a YAWNING-TITAN~\cite{andrew2022developing} environment modified to allow attackers and defenders to act as independent learning agents.
We leverage the proximal policy optimization reinforcement learning algorithm of Schulman \textit{et al.}~\cite{schulman2017proximal}, an on-policy deep reinforcement learning algorithm with generally good performance that is used in the default implementation of YAWNING-TITAN\footnote{https://github.com/dstl/YAWNING-TITAN/tree/main}.
We conclude with a discussion of our results and avenues for further work.
\subsection*{Contributions}
This work evaluates the ``price of pessimism'', a phenomenon wherein the \textit{a priori} assumption about an adversary's knowledge of a system results in a suboptimal response pattern.
In particular, this manuscript contributes the following:
\begin{enumerate}
    \item For reinforcement learning agents in a stochastic Bayesian game, optimizing against a worst-case adversary leads to suboptimal policy convergence.
    \item Defending agents trained against attacking agents that learn are also highly capable against algorithmic attackers even when they have not seen those algorithmic attackers during training.
    \item An extension of the YAWNING-TITAN~\cite{andrew2022developing} reinforcement learning framework for training independent attacking and defending agents
    \item A novel use of the Bayes-Hurwicz criterion for parameterizing attacker decision making under uncertainty 
\end{enumerate}
\section{Related Work}
Security game theory is a broad field informed by cybersecurity, decision theory, and game theory.
Recent challenges like CAGE~\cite{ttcp2021cage} have encouraged development of models like CybORG~\cite{foley2022autonomous} and YAWNING-TITAN~\cite{andrew2022developing} that use reinforcement learning to train autonomous agents that defend against cyber attacks.
The Ph.D thesis of Campbell~\cite{campbell2022autonomous} also considers a similar problem space to our work and leverages the same game theoretic model.
These works address a similar problem space to our work: the development of a defensive agent that disrupts an adversary while minimizing impact to network users.
This paper builds on prior work by the authors~\cite{galinkin2023simulation} that uses a simple state and action space for Stochastic Bayesian Games (SBG) as introduced by Albrecht and Ramamoorthy~\cite{albrecht2013game}.
The partial observability of the proposed SBG relates closely to the work of Tom\'{a}\v{s}ek, Bo\v{s}ansk\'{y}, and Nguyen~\cite{tomavsek2020using} on one-sided partially observable stochastic games. 
Their work considers sequential attacks on some target and develops scalable algorithms for solving zero-sum security games in this setting and present algorithms to compute upper and lower value bounds on a subgame.
By contrast, our work seeks to understand how the defender's beliefs about the attacker impacts the rewards and outcomes for defenders. 
Additionally, the aforementioned works and other related works like Khouzani \textit{et al.}~\cite{Khouzani2012Saddle-pointAttack} and Chatterjee \textit{et al.}~\cite{Chatterjee2016PropagatingAnalysis} consider the attacker as either a deterministic operator or leverage epidemic modeling techniques to describe an attacker's movements through a network.
Our work here is unique in the respect that we model the attacker as a learning agent, a phenomenon more in line with real-world attackers.

The work of Thakoor~\textit{et al.}~\cite{thakoor2020exploiting} and subsequent work by Aggarwal~\textit{et al.}~\cite{aggarwal2022designing} informs our point of view on how attackers respond to risk.
In this work, we implicitly assume bounded rationality and account for risk and uncertainty within our model.
While Thakoor~\textit{et al.} and Aggarwal~\textit{et al.} use cumulative prospect theory~\cite{tversky1992advances} to address deception as a source of uncertainty, we instead consider it one component of a larger overall framework.

A key component of this work is the assumptions about information available to the players.
Specifically, the defending player's beliefs about an attacker's knowledge.
In the security games context, this dynamic is well-captured by existing literature on deception and counter-deception~\cite{pawlick2021game}.
Our work extends this research area by exploring this dynamic from a game design perspective concerning beliefs defending players hold about the attackers independent of in-game actions and how that impacts the learning of defensive agents.
\section{Modeling Assumptions} \label{sec:assumptions}
Since our work is concerned with modeling assumptions, we aim to make our own assumptions as explicit and general as possible.
We extend the state and action space used in prior work~\cite{galinkin2023simulation}, but operate under the same assumption that attackers and defenders choose their next action simultaneously at each time step.
We maintain, without loss of generality, that there is a single attacker and a single defender present in the game.

\subsection{State Space}
The state space $S$ of this game consists of a network graph $G = (V, E)$, where each $v \in V$ is a defender-owned computer and each edge $e \in E$ is a tuple indicating a network connection between two nodes $(u, v); u, v \in V$. 
The state of each machine $v \in V$ is a tuple $(v_p, v_{\alpha}, v_{\delta})$.
\begin{itemize}
    \item $v_p \in [0, 1]$ is the ``vulnerability'' of a node: the probability that a basic attack will be successful
    \item $v_{\alpha} = \{0, 1\}$ is the ``true'' state of compromise and is visible only to the attacker
    \item $v_{\delta} = \{0, 1\}$ is the defender-visible state of compromise
\end{itemize}
At each time step, with probability $q$, an ``alert'' is generated independent of attacker or defender action that sets $v_{\delta} = 1$ even if the true compromise state $v_{\alpha} = 0$, corresponding to a false positive alert.
This phenomenon is justified and described in further depth in Section~\ref{subsec:noise}.

\subsection{Attackers} \label{sec:attackers}
The attacker's action space, $A_{\alpha}$ consists of actions on elements of $V$ subject to visibility constraints.
We define a ``compromised'' node as a node that the attacker has gained access to and thus has $v_{\alpha} = 1$.
An ``accessible'' node is any node with some edge connecting it to any compromised node.
The observable state space of an attacker, $S_{\alpha}$ consists of all compromised nodes and all accessible nodes.
The attacker's action space consists of the following actions, which each incur some cost $c$:
\begin{itemize}
    \item Basic Attack: Compromise an accessible $v \in V$ with probability $v_p$.
    \item Zero-day Attack: Compromise an accessible $v \in V$ as if $v_p = 1$.
    \item Move: Move from some compromised $v \in V$ to another compromised $v' \in V$
    \item Do Nothing: Take no action
    \item Execute: End the game and realize rewards for all compromised $v \in V$
\end{itemize}

Attacker types inform what exactly their objectives are and may influence the categories of malware used \textit{e.g.}, coin miner, ransomware, backdoor.
Moreover, the attacker's type informs the utility function of the attacker and the cost of each action. 
For some attacker types, \textit{e.g.}, cybercriminals using ransomware, there is a clear utility: the ransom paid by the victim.
However, other attacker types may aim to steal private information that is not be directly convertible to currency.
The estimation of these utility functions is thus type-specific, though any compromised machine will confer nonzero utility to the attacker.

For our purposes, we assume that the attacker is a ransomware attacker and aims to compromise as many machines as possible and end the game before the defender can remove them from the system.
Assuming unit reward for each node, this means that for a network of size $n$, the attacker's reward at any time if they take the Execute action or control an attacker-defined percentage of the network is:
\[u_{\alpha} = \sum_{i = 0}^{n} v_{i\alpha} - \sum_{t = 0}^{T} c_{t}\]
Where $T$ is the final timestep of the game and $c_{t}$ is the cost of the action taken at time $t$.
We note that our reward function in Section~\ref{sec:empirical} uses a scaling factor for the value of $v_{i \alpha}$ in lieu of unit value, and that this function also holds in cases where each node has a different value.
\subsection{Defenders} \label{sec:defenders}
The defender's action space, $A_{\delta}$ consists of actions on $V$ and $E$.
Although the defender can take only one action at each time step, they may take that action on a set of nodes or edges.
The defender's observable space, $S_{\delta}$ consists of the entirety of $E$ and the number of alerts, $v_{delta}$, for all $v \in V$ at all times.
Specifically, the defender may:
\begin{itemize}
    \item Reduce Vulnerability: For some $v \in V$, slightly decrease the probability that a basic attack will be successful
    \item Make Node Safe: For some $v \in V$, reduce the probability that a basic attack will be successful to 0.01
    \item Restore Node: For some $v \in V$, reset the node to its initial, uncompromised state, including the probability that a basic attack will be successful
    \item Scan: With some probability, detect the true compromised status of each $v \in V$
    \item Do Nothing: Take no action
\end{itemize}

The objective of cybersecurity, broadly, is to maintain the confidentiality, availability, and integrity~\cite{anderson2020security} of a system. 
The defender's utility thus arises from the availability of resources.
Each action has some cost $c$ associated with it, where the impact of the action being taken on the availability of that resource on the system dictates $c$.
For example, the Reduce Vulnerability and Make Node Safe actions are very similar, but the Reduce Vulnerability action incurs a much smaller cost under default YAWNING-TITAN settings.
The defender's utility $u_{\delta}$ is a fixed-value reward for eliminating the adversary or withstanding the attack minus the sum of all costs associated with the actions taken during the course of the game.
\subsection{Presence of Noise} \label{subsec:noise}
Security detections are not infallible, and some number of both false positives -- detections that alert on benign behavior, and false negatives -- failures to detect malicious behavior, must be expected. 
Attackers are incentivized to  and have adopted techniques like using cloud infrastructure and software as a service (SaaS) providers to conduct attacks~\cite{Galinkin2019TheCloud} and the use of legitimate executables or ``lolbins'' for malicious purposes ~\cite{kumar2020emerging}. 
Attackers seek to blend in, so detection of malicious behavior that is similar to benign behavior is important for defenders.
Since there is no way to definitively determine whether or not a program is malicious, these rules and algorithms yield some number of false positive alerts.
The empirical rate of these false positive alerts, according to surveys, appears to be somewhere between 20\%~\cite{orca2022cloud} and 32\%~\cite{kohgadai2017alert}. 
In cases where some number of alerts are not an indicator of actual attack activity, any probabilistic approach to network security must grapple with this noise.
The security game setting has modeled this sort of behavior in the realm of deception and counter-deception. 
Work by Nguyen and Yadav~\cite{nguyen2022risk} shows that while attacker payoffs are improved by deception, learning defenders can reduce the value of this deception.

Assuming that an attacker's behavior is detected with some probability $p$, then there is an independent probability of false positives $q$.
Letting $p$ and $q$ characterize two independent Bernoulli processes that may each yield an alert, we can treat the emission of an alert as the joint probability of these two processes.
The probability of an alert occurring at all is thus $p + q - pq$, as described in earlier work by the authors~\cite{galinkin2023simulation}.
The expected probability that a particular alert is attributable to benign activity is $(1-p) q$.
For simplicity, we assume that $p$ and $q$ are the same across all nodes.

In the absence of the noise assumption, attacker-defender interaction becomes a game of Cops and Robbers on a graph~\cite{simard2021general} where a defender can eliminate the attacker by finding the ``cop number'' -- the number of nodes required to ``surround'' the attacker and eliminate all of their access at once -- for the subgraph the attacker has explored.
This is still an extremely challenging problem, since even without noise, the defender only has a belief about the extremal edges of that subgraph and finding the attacker's possible subgraphs has exponential complexity. 
As a result, the importance of an assumption about noise relates with assumptions about under what circumstances a defender realizes a reward.
\subsection{Presence of Attackers} \label{presence}
In non-cooperative game theory, two players are playing a game and each seeks to optimize against some utility function.
This comes, of course, with the implicit assumption that both players know they are playing a game.
In cybersecurity games, when modeling the beliefs of the defender, we frequently imply that the defender knows an attacker is present \textit{a priori}. 
In reality, attackers are not always present in our system, and this has a substantial impact on defender expectations.
Clearly, any response taken when an attacker is not present incurs a cost and yields no reward. 

Let $\mu \in [0, 1]$ be the probability that an attacker is present in a system.
If our game has alert probability $p$ and no noise -- that is, $q = 0$ -- then although we know the probability of an alert is $\mu p$, the occurrence of an alert allows us to set $\mu = 1$ and the defender can directly pursue the attacker as described in Section~\ref{subsec:noise}.
Assuming noise is present in the system, the probability of an alert occurring at any time step is then
\begin{equation} \label{eqn:alert_probability}
    (1 - \mu) q + \mu (p + q - pq)
\end{equation}

In this setting, since an attacker may not be present, the probability of a false positive event occurring at any time step is
\[(1 - \mu) q + \mu (1-p) q\]
\section{Attacker Knowledge}
As one might predict and as was demonstrated in prior work~\cite{galinkin2023simulation}, attacker utility increases monotonically with the knowledge available to them.
Knowing the parameters of a simplified game allows defenders to set a threshold of alerts in expectation -- that is, if they know the values of $p$, $q$, and $\mu$, they can compute the number of alerts that might be expected at time $t$ and shut down any system that has generated more than the threshold number of alerts.
However, considering only the value in expectation can lead to poor outcomes for the defender, as even for small values of $q$, any deviation from expectation can lead to suboptimal outcomes for the defender \textit{e.g.} shutting down systems when no attacker is present.
As such, the defender's threshold for taking an action should instead be influenced by the level of deviation from their expectation.

As in Section~\ref{presence}, our per-node expectation of alerts, conditioned on the presence of an attacker, is $(1 - \mu) q + \mu(p + q - pq)$.
The defender establishes some prior $\mu$ and at each time step $t$, observes some number of alerts across the $n$ nodes of the network, where $n = \left| V\right|$.
The total number of alerts, $Al$, expected at time $t$ is therefore $nt ((1-\mu) q + \mu(p + q - pq))$, which can be treated as a random sample drawn from a Beta-binomial distribution.
Given $Al$, the defender performs a Bayesian update on $\mu$. 
The posterior distribution of $\mu$ is a beta distribution and so at time $t$, the defender updates $\mu$ as follows:
\begin{align} \label{eqn:mu_update}
    \mu_t & = \int_{0}^{1} \frac{\mu_{t-1}^{\alpha - 1}(1 - \mu_{t-1})^{\beta - 1}}{B(\alpha, \beta)}
\end{align}
Where $\phi = p + q - pq$ for brevity, $B$ is the beta function, and $\alpha, \beta$ are empirically derived parameters such that:
\[Al = \frac{\alpha}{\alpha + \beta}\]
\[Var[Al] = \frac{\alpha \beta}{(\alpha + \beta)^2 (\alpha + \beta + 1)}\]
This yields the following for deriving $\alpha$ and $\beta$ from observed alerts:
\[\alpha = \left(\frac{1 - Al}{Var[Al]} - \frac{1}{Al}\right) Al^2\]
\[\beta = \alpha\left(\frac{1}{Al} - 1\right)\]

Setting $\mu = 0$, the defender should observe, on average, $nq$ alerts at each time step.
As the defender observes more than $nq$ alerts at each time step, their confidence that an attacker is present grows. 
In lieu of simply setting an alert threshold, the defender sets some threshold for $\mu$ according to their risk tolerance, which may be calculated given the value of the network compared to costs or given exogenously. 
Once this empirical posterior estimate of $\mu$ is exceeded, remediation action should be taken.

In the full-knowledge scenario, however, the attacker has access to all of the defender's information and can see the threshold for $\mu$.
Furthermore, they can directly compute how an action they will take will update $\mu$ and choose an action in accordance with that update, subject only to the condition that due to the presence of noise, some triggering event may occur even if they take no action.
Thus, just as in prior work~\cite{galinkin2023simulation}, the attacker should take the Execute action to end the game and collect the current reward when taking any other action may push $\mu$ over the threshold.
\subsection{Zero Knowledge} \label{sec:zero_knowledge}
In real-world settings, attackers are unaware of parameters like $p$ and $q$, so they cannot \textit{ex ante} optimize their actions and must instead seek to achieve their goal by balancing the risk of being caught with the need to achieve their goal.
Thus, the ``optimistic'', from the defender's perspective, zero knowledge setting is the most consequential for generating real security impacts.
In the zero knowledge setting, the defender is still armed with full knowledge and chooses some threshold for $\mu$ to take an action.
However, the attacker does not know any of the parameters of the network and must balance exploration and exploitation given only the state, as each action they take may alert the defender of their presence.
Given their limited information, they can use the Restricted Bayes/Hurwicz criterion~\cite{ellsberg2015risk} to choose their action.

\begin{definition}
    The Restricted Bayes/Hurwicz criterion is a procedure for decision making under uncertainty parameterized by $\gamma$ and $delta$ defined by:
    \begin{equation} \label{eqn:bh}
        \gamma \hat{P} + (1 - \gamma)[\delta \bar{P}_{x^*} + (1-\delta) \ubar{P}_{x^*}]
    \end{equation}
    where $\gamma$ is the decision-maker's confidence in their distribution, $\delta$ is the coefficient of pessimism, $\hat{P}$ is the prior distribution over the decision-maker's actions, $\bar{P}_{x^*}$ is the best case probability distribution over actions, and $\ubar{P}_{x^*}$ is the worst case probability distribution over actions
\end{definition}

The attacker has some initial probability distribution $\hat{P}$ over their actions indicating the probability they will take an action at a given time step given a state.
The coefficient of pessimism $\delta$ corresponds to the attacker's belief about the value of $\mu$, which we write $\hat{\mu}$.
Due to the limited signal available to the attacker, the confidence parameter $\gamma$ should be monotonically decreasing over time as they observe likely defender actions.
For notational simplicity, we set $\delta = 1 - \hat{\mu}$ and write the criterion as follows:
\[\gamma \hat{P} + (1 - \gamma)[(1 - \hat{\mu}) \bar{P}_{x^*} + \hat{\mu} \ubar{P}_{x^*}]\]

At the start of the game, the attacker must estimate the probability of an alert occurring when they take an action. 
The attacker knows that false positive alerts occur and can infer that they need to establish some value that reflects Equation~\ref{eqn:alert_probability}.
In the absence of any information about the defender's configuration, the attacker must sample some value $\hat{\phi} \in [0, 1]$.
The use of the PERT distribution~\cite{clark1962pert}, a special case of the Beta distribution, is well-motivated from operations research.
We allow the minimum and maximum value in the range $[0, 1]$ and set the $b$ parameter of the distribution to the midpoint $0.5$.
% The distribution of $\hat{\phi}$ is shown in Figure~\ref{fig:pert}.
The initial value of $\hat{\mu}$ can be similarly sampled.

% \begin{figure}[h]
%     \centering
%     \includegraphics[width=0.6\textwidth]{pert.png}
%     \caption{Distribution for $\hat{\phi}$ and $\hat{\mu}$: A PERT distribution with $a=0, b=0.5, c=1$}
%     \label{fig:pert}
% \end{figure}

When the attacker observes a signal from a defender -- that is, when the defender takes some action on an attacker-controlled node, the attacker updates their belief about $\hat{\mu}$, and reduces $\gamma$, since they are less confident in their initial distribution over their actions.
This update process follows Algorithm~\ref{alg:update}.
In order to conduct this update, the attacker estimated alert generation probability $\hat{\phi}$ is used alongside their estimate of the defender's belief about an attacker's presence, $\hat{\mu}$ since the attacker has no knowledge of the number of alerts and must instead construct a Bayesian estimate $\hat{Al}$ given these parameters.

\begin{algorithm}
\caption{Attacker Parameter Update Algorithm}\label{alg:update}
\begin{algorithmic}
\Require $\hat{\mu} \in [0, 1]$, $\hat{\phi} \in [0, 1]$
\State $k \gets 0$ \Comment{$k$ is the number of observed defender actions}
\State $\gamma \gets 1$
\State $\hat{Al} \gets nt\hat{\mu}\hat{\phi}$
\State $Var[\hat{Al}] \gets nt\hat{\mu}\hat{\phi}(1-\hat{\phi})$
\While{not done}
\If{defender action observed}
    \State $k \gets k + 1$
    \State $\alpha = \left(\frac{1 - \hat{Al}}{Var[\hat{Al}]} - \frac{1}{\hat{Al}}\right) \hat{Al}^2$
    \State $\beta = \alpha\left(\frac{1}{\hat{Al}} - 1\right)$
    \State $\gamma = \gamma_k / (k + 1)$
    \State $\hat{\mu} = \int_{0}^{1} \frac{\hat{\mu}_{k}^{\alpha - 1} (1 - \hat{\mu}_{k})^{\beta - 1}}{B(\alpha, \beta)}$
\EndIf
\EndWhile
\end{algorithmic}
\end{algorithm}

In the full-knowledge case with thresholding, the best response dynamics are determined exactly by expectation and the parameters of the network~\cite{galinkin2023simulation}.
However, this work considers a significantly expanded state space where both attackers and defenders have a richer action space.
In both the full-knowledge and zero-knowledge case, the attacker's choice of action depends on the state and any actions observed from the defender.
The zero-knowledge case in particular is highly dependent on the establishment of good prior distributions. 
Prior distributions can be given exogenously or can be learned.
In our case, we elect to learn $\hat{P}, \bar{P}, \ubar{P}$ via simulation.
The results of this simulation are described in Section~\ref{sec:priors}.
\section{Empirical Evaluation} \label{sec:empirical}
Game theoretic proofs about behavior in cybersecurity environments can provide powerful tools for thinking about how attacker-defender interaction occurs in practice.
However, they do not always carry over to real-world environments.
To gain empirical insight into the way these assumptions manifest in practice, we modify the YAWNING-TITAN (YT) framework~\cite{andrew2022developing} to include noise and allow two independent agents -- one attacker and one defender -- to be trained simultaneously.
To do this, we create a new multiagent environment with a single state space where an attacking agent and a defending agent both operate but have their own separate observation and action spaces.
This environment extends the functionality of YT by providing the ability to treat the attacking agent as a learner, rather than following a fixed algorithm for determining attacker actions.
Moreover, since our zero-knowledge training case involves training both an attacker and a defender with different observation spaces, methods like Multi-Agent DDPG~\cite{lowe2017multi} are not suitable, as such methods would disclose hidden information about the environment to each agent.
Despite issues of known overfitting to suboptimal policies due to non-stationarity~\cite{moalla2024no}, we opt to use two distinct instances of proximal policy optimization (PPO)~\cite{schulman2017proximal} -- one each for the attacker and defender -- to train our agents, as this the algorithm generally performs well on a variety of tasks and has been used in prior, related work by others~\cite{andrew2022developing}.

In this environment, we associate a cost to each attacker and defender action in accordance with those included in YT, and associate a positive reward for the agent that ``wins'' the episode along with a negative reward for the agent that ``loses'' the episode.
To improve learning, we scale the negative reward for the defender such that they achieve a lesser negative reward for closer failures.
Specifically, this scaling factor is the number of timesteps that the game has taken divided by the maximum number of timesteps required for a defender victory.
We find that our defensive agent trained to expect the adversary has complete knowledge of the system will make worse decisions by overestimating the adversary when less information is available. 
For the sake of evaluating defensive agents against a programmatic, non-learning attacker, we adapt YT's \texttt{NSARed} agent~\cite{ridley2018machine}, based on a description of cyber attack automation, to work within our environment.
The implementation of the \texttt{NSARed} agent is purely algorithmic -- there is no machine learning component -- and defensive agents are not exposed to the agent at training time, making it a stable baseline for comparison.
The full code for our training and evaluation implementation is available on GitHub\footnote{https://github.com/erickgalinkin/pop\_rocks/}.

\subsection{Establishment of Priors} \label{sec:priors}
The attacker must consider three distributions when playing the game: $\hat{P}, \bar{P}, \ubar{P}$.
For the ``best case'' distribution $\bar{P}$, we train an attacking agent against a defender whose only action is to do nothing.
For the ``worst case'' distribution $\ubar{P}$, we train an attacking agent against a defender who has access to both attacker and defender observation spaces and thus has full-knowledge of the attacker's moves and the network.
The remainder of this subsection describes the training of the model $\hat{P}$ is drawn from.

Our setting assumes an adversary who uses ransomware, where the attacking player ``wins'' when they control more than 80\% of the network.
For each training episode, we instantiate a random entrypoint for the attacker on a 50-node network whose edges are randomly generated to ensure the network has 60\% connectivity, and that there are no unconnected nodes.
We leverage proximal policy optimization (PPO)~\cite{schulman2017proximal} for our learning agents in two settings:
\begin{enumerate}
    \item Optimistic (zero-knowledge): The attacking agent can see only the nodes they control and adjacent nodes. They cannot see the vulnerability status of any nodes.
    \item Pessimistic (full-knowledge): the attacking agent has access to the same information and observation space as the defending agent.
\end{enumerate}
In accordance with our modeling assumptions, the attacking and defending player simultaneously decide their moves for timestep $t$ from their action space.
Each agent is trained in either the optimistic or pessimistic environment for 3000 episodes and evaluated for 500 episodes across randomly generated environments in both the optimistic and pessimistic setting.
Based on empirical results from experiments in the environment, the actor learning rate is set to 0.0002 and the critic learning rate is set to 0.0005. 
Higher learning rates were tried and led to fast convergence to suboptimal policies, as PPO assumes full observability to achieve globally optimal policies and our environment is only partially observable.
Values for all hyperparameters and action costs for both players were fixed across all settings and are included in Tables~\ref{tab:params} and \ref{tab:costs}.

\begin{table}
\parbox{.45\linewidth}{
\centering
\begin{tabular}{|l|l|}
\hline
\textbf{Parameter} & \textbf{Value} \\
\hline
Actor Learning Rate           & 0.0002          \\
Critic Learning Rate          & 0.0005          \\
Training Epochs               & 3500            \\
$\gamma$                      & 0.99          \\
Update Epochs                 & 5               \\
Batch Size                    & 64              \\
Win Reward                    & 5000        \\
Lose Reward                   & -100  \\
\hline
\end{tabular}
    \caption{Hyperparameter values for reinforcement learning experiments.}
    \label{tab:params}
}
\parbox{.45\linewidth}{
\centering
\begin{tabular}{|l|l||l|l|}
\hline
\multicolumn{1}{|c|}{\textbf{Attacker}} & \multicolumn{1}{|c||}{\textbf{Cost}} & \multicolumn{1}{|c|}{\textbf{Defender}} & \multicolumn{1}{|c|}{\textbf{Cost}} \\
\hline
Basic                        & 2                                 & Reduce Vuln                & 1.5                               \\
Zero Day                            & 6                                 & Make Safe                      & 4                                 \\
Move                                & 0.5                               & Restore                        & 6                                 \\
Do Nothing                          & 0                                 & Scan                                & 0.5                               \\
Execute                             & 0                                 & Do Nothing                          & 0                    \\
\hline
\end{tabular}
    \caption{Action costs for attackers and defenders}
    \label{tab:costs}
}
\end{table}

\begin{figure}[!tbp]
  \centering
  \begin{minipage}[b]{0.45\textwidth}
    \includegraphics[width=\textwidth]{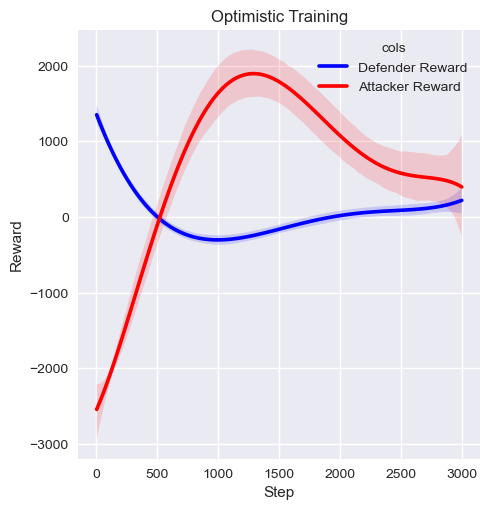}
    \caption{Training reward curves for attacking and defending agents in the optimistic setting}
    \label{fig:opt-train}
  \end{minipage}
  \hfill
  \begin{minipage}[b]{0.45\textwidth}
    \includegraphics[width=\textwidth]{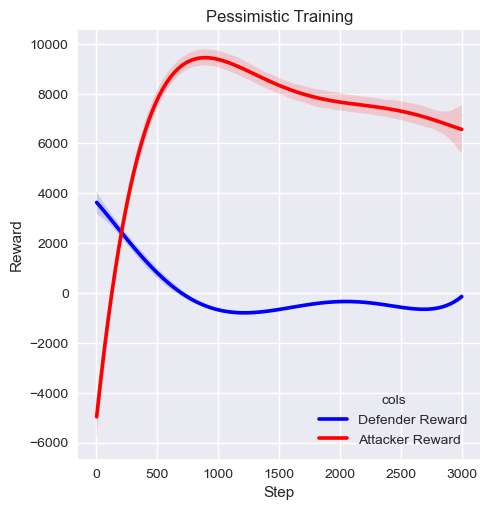}
    \caption{Training reward curves for attacking and defending agents in the pessimistic setting}
    \label{fig:pess-train}
  \end{minipage}
\end{figure}

\begin{figure}
    \centering
    \includegraphics[width=0.7\textwidth]{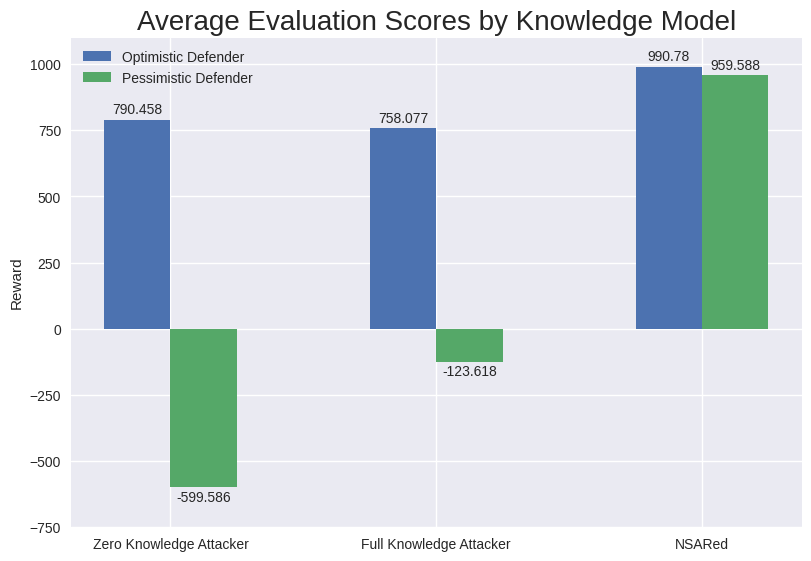}
    \caption{Average reward for optimistic and pessimistic trained defending agents across 500 evaluation trials against zero knowledge, full knowledge, and \texttt{NSARed} attackers.}
    \label{fig:knowledge}
\end{figure}

\begin{figure}
    \centering
    \includegraphics[width=0.7\textwidth]{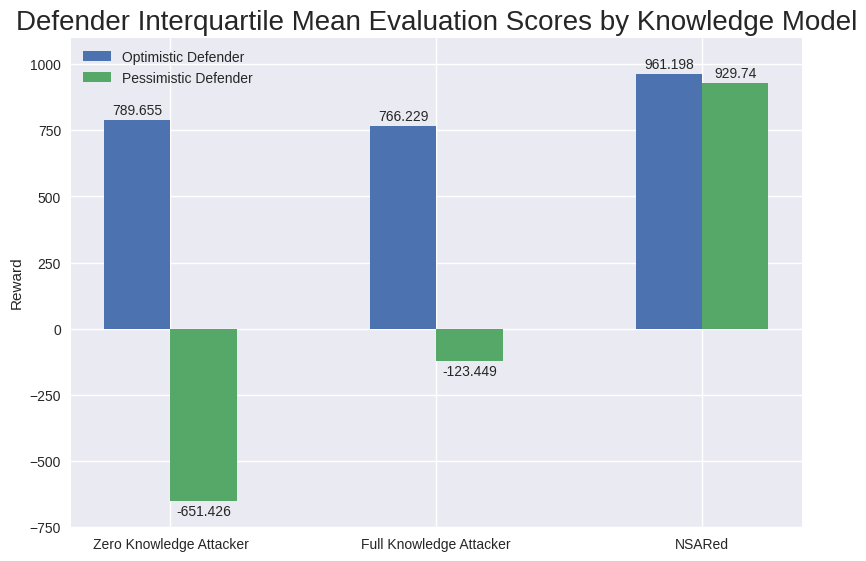}
    \caption{Interquartile mean reward for optimistic and pessimistic trained defenders across 500 evaluation trials against zero knowledge, full knowledge, and and \texttt{NSARed} attackers.}
    \label{fig:iqm}
\end{figure}

\begin{figure}[h!]
    \centering
    \includegraphics[width=0.8\textwidth]{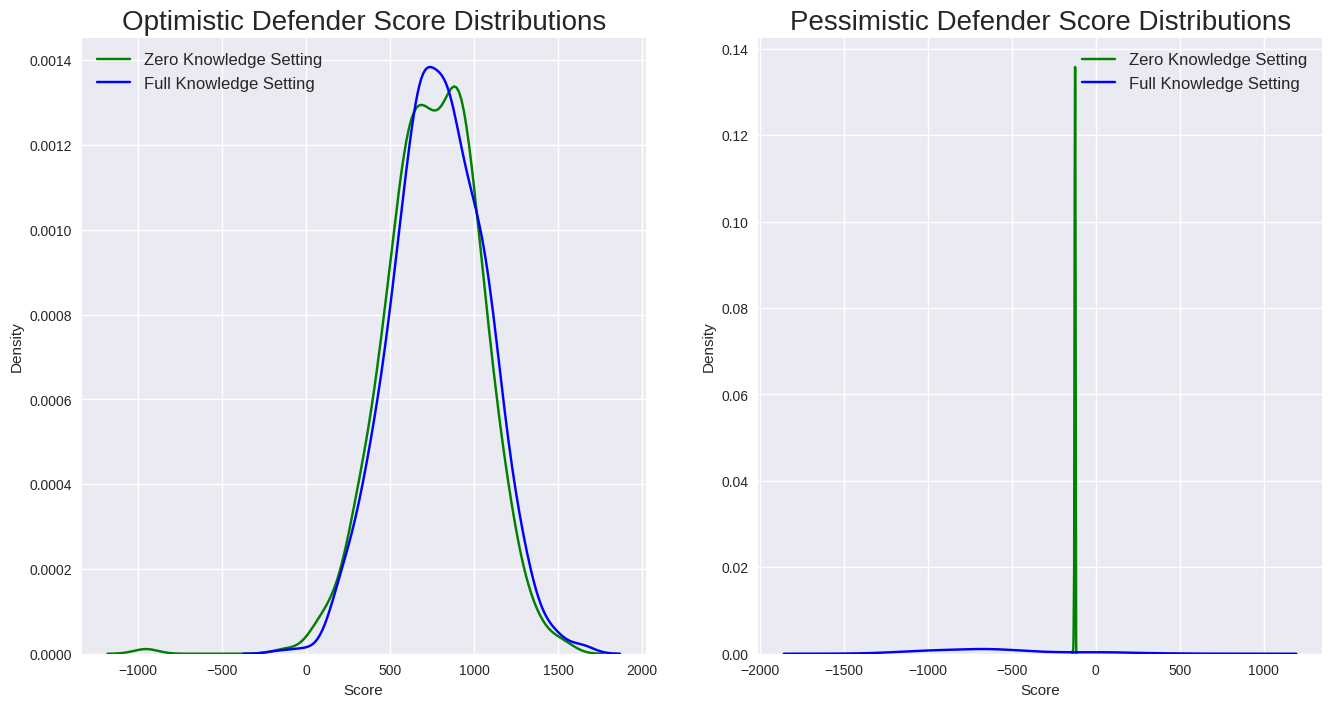}
    \caption{Score Distributions for optimistic and pessimistic trained defenders across 500 evaluation trials in zero knowledge and full knowledge settings. Note that the scale of x and y axes differ between the subfigures.}
    \label{fig:sd}
\end{figure}

Evaluating reinforcement learning findings is notoriously difficult.
Therefore, in line with Agarwal \textit{et al.}~\cite{agarwal2021deep}, we look to more robust measurements that capture the uncertainty in results.
Specifically, in addition to standard evaluation metrics, we consider also score distributions and the interquartile mean across evaluation runs. 
These metrics help capture the stochasticity in the task and normalize our results.
Smoothed training curves in the optimistic setting are shown in Figure~\ref{fig:opt-train}; the pessimistic setting is shown in Figure~\ref{fig:pess-train}.
The average rewards and interquartile mean for evaluation of the defending agents trained in the optimistic and pessimistic settings achieved each of the two evaluation settings are shown in Figure~\ref{fig:knowledge} and Figure~\ref{fig:iqm}.
Score distributions are captured in Figure~\ref{fig:sd}.

We observe in Figures~\ref{fig:opt-train} and \ref{fig:pess-train} that the attacker generally starts out with a poor reward, but learns how to attack the target within 500 epochs.
In the optimistic setting, the defender initially starts out with a very high level of reward but once the attacker begins winning more often, they must adapt their strategy, with rewards for both agents converging around epoch 3000.
In the pessimistic setting, the defender similarly starts with a reasonably high reward, but the attacker quickly learns how to overcome the defender's strategy. 
In this setting, the defender does not rebound and instead settles into a local optimum -- minimizing the magnitude of loss rather than continuing to explore for a strategy that eliminates the attacker.
Longer runs -- up to 10000 training epochs -- were attempted, but the pessimistic defender's policy in those cases achieved even worse evaluation results than the defender trained for 3000 epochs.
Note that the reward scale for attackers and defenders is not the same and defenders cannot achieve the extreme rewards that attackers do.

What we find from our evaluation is that the defending agent trained in the pessimistic setting performs worse on average than the defending agent trained in the optimistic setting.
Across both evaluation settings, the optimistic defender is more robust in general and performs significantly better on average across all settings, as we can observe from Figures~\ref{fig:knowledge} and \ref{fig:iqm}.
Note that a negative score implies that the attacker is winning more often than the defender, while a positive score implies that the defender is winning more often.
Each defender performs relatively better in the setting they were trained in and experiences less variance, as we can see the relative stability between the mean and interquartile mean for in-domain settings.
Interestingly, both optimistic and pessimistic defenders perform well against the \texttt{NSARed} attacker, suggesting that training against learning agents offers substantial benefit over training against more ``static'' algorithmic attackers.

The score distributions in Figure~\ref{fig:sd} demonstrate the impact of the pessimistic defender's convergence. 
While the optimistic defender has a fairly broad, relatively normal distribution in both settings, the pessimistic defender's probability density is highly concentrated just below zero in domain and widely distributed across highly negative and highly positive out of distribution. 
This underscores the robustness of the optimistic defender and indicates that in training, the pessimistic defender converges to a policy that expects to achieve a negative reward and seeks to minimize that negative reward, rather than a policy that expects a positive reward and seeks to maximize it.
Since the defender's win condition depends on eliminating the attacker or surviving 500 episodes, this ``loss minimization'' policy leads to suboptimal performance against learning attackers.

To explain the difference in outcomes and what is learned in training, we can examine the differences in actions taken by the optimistic and pessimistic defenders.
We find that the pessimistic defender agent uses the more expensive ``restore node'' action at a higher frequency than the optimistic agent, while the optimistic agent spends more turns on reducing the vulnerability of nodes and making nodes safe. 
This is likely an artifact of the attacker in the pessimistic setting having access to significantly more information, requiring a more aggressive response to forestall an attacker win.
The distribution of action usage for both agents across all evaluations is shown in Table~\ref{tab:actions}.

\begin{table}
\centering
\resizebox{0.7\textwidth}{!}{%
\begin{tabular}{|l|l|l|l|}
\hline
\textbf{Action}               & \textbf{Pessimistic} & \textbf{Optimistic} & \textbf{Difference}\\
\hline
\textbf{Do Nothing}           & 0\%            & 1.4\%         & 1.4\% \\
\textbf{Scan}                 & 3.22\%         & 3.19\%        & 0.03\% \\
\textbf{Reduce Vulnerability} & 16.1\%         & 20.68\%       & 4.58\% \\
\textbf{Make Node Safe}       & 38.83\%        & 41.33\%       & 2.5\% \\
\textbf{Restore Node}         & 41.85\%        & 33.4\%        & 8.45\% \\
\hline
\end{tabular}%
}
    \caption{Percentage of Actions Taken by Defending Agent Across all evaluation episodes, ordered in ascending cost of the action.}
    \label{tab:actions}
\end{table}

In the interest of ensuring our action costs do not have an undue influence on our results, we perform our same reinforcement learning evaluation for our pessimistic setting.
We vary two parameters -- ``connectivity'' and ``security'' -- over the range $(0,1]$, with a floor of 0.000001 to avoid division by zero.
To establish the robustness of the model, we multiply the cost of each action by the ratio of connectivity and security such that the importance of security is nearly zero, the cost of actions becomes very high and when the value of connectivity is nearly zero, the cost of actions approaches zero, given fixed security value. 
We find that there is a nearly linear pattern relating the cost of actions to rewards, but the learned policy remains stable, provided the \emph{relative} costs of actions is fixed and is not degenerate even near extremal values.
% We plot these results in Figure~\ref{fig:ablation}, omitting values where the connectivity parameter is near zero for readability, as the defending player always achieves an extremely high reward.

% \begin{figure}[h!]
%     \centering
%     \includegraphics[width=0.6\textwidth]{ablation.png}
%     \caption{Average reward for defending player across 500 evaluation trials for varying ``connectivity'' and ``security'' settings.}
%     \label{fig:ablation}
% \end{figure}

\subsection{Use of Bayes-Hurwicz Decision Criterion for Attackers}
As mentioned in Section~\ref{sec:zero_knowledge}, attackers in the real world do not have the luxury of training their priors to convergence against a target and must combine their prior knowledge with what is observed during an attack.
Since the attacker is making decisions under uncertainty, some criterion must be used to allow them to do that subject to their own parameters.
Using the pretrained, frozen models from Section~\ref{sec:priors}, we consider how the application of the Restricted Bayes/Hurwicz criterion for the attacker as defined in Equation~\ref{eqn:bh} impacts the outcomes of the attacking player.
Aside from the use of the Restricted Bayes/Hurwicz criterion for the attackers, all of the defender and environmental evaluation settings remain the same.

The attacking agent draws independent prior $\hat{\mu}$ and $\hat{\phi}$ values from PERT distributions in accordance with Algorithm~\ref{alg:update} at the start of each round and updates their $\hat{\mu}$ at each timestep if defender activity is observed -- that is, if the defender takes an action that removes access to an attacker controlled node.
For each evaluation scenario, the relevant trained attacker model -- zero knowledge or full knowledge -- is used to establish $\hat{P}$, $\bar{P}$, and $\ubar{P}$ for the observed state of the game. 
Specifically, each policy model $\hat{\pi}, \bar{\pi}, \ubar{\pi}$ accepts an attacker-observed state $S_t$ and outputs a distribution over the attacker's action space $A_{\alpha}$ such that $\hat{\pi}(S_t) \sim \hat{P}$, $\bar{\pi}(S_t) \sim \bar{P}$, and $\ubar{\pi}(S_t) \sim \ubar{P}$.
The attacker leverages these model outputs and the values of $\hat{\mu}$, and $\gamma$ with Equation~\ref{eqn:bh} to determine their next best action.

\begin{figure}[h!]
    \centering
    \includegraphics[width=0.7\textwidth]{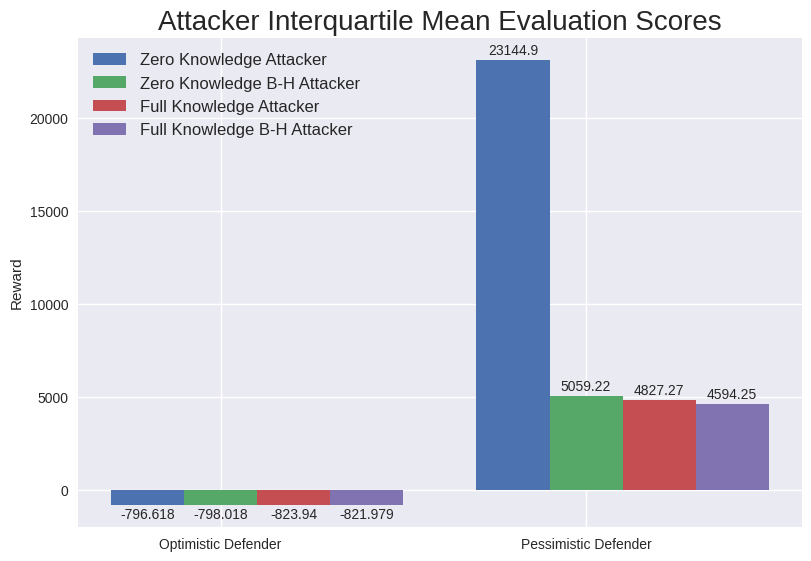}
    \caption{Interquartile mean reward for attacking players across 500 evaluation trials against both optimistic and pessimistic defenders}
    \label{fig:iqm_red}
\end{figure}

\begin{figure}[h!]
    \centering
    \includegraphics[width=0.7\textwidth]{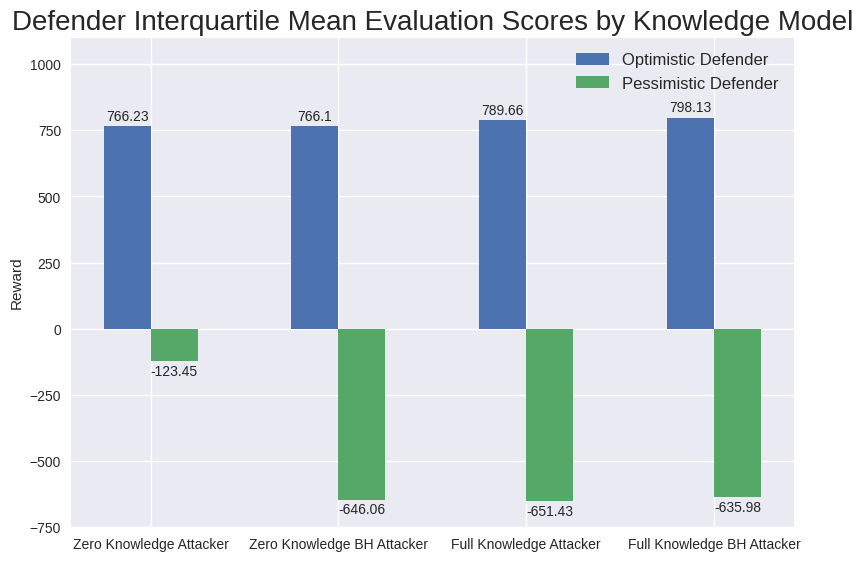}
    \caption{Interquartile mean reward for optimistic and pessimistic defenders across 500 evaluation trials against all attacker types.}
    \label{fig:bh_blue}
\end{figure}

Interquartile mean evaluation rewards for the attacker, shown in Figure~\ref{fig:iqm_red}, illustrate the impact of restricted Bayes/Hurwicz. 
Against the optimistic defender, the use of restricted Bayes/Hurwicz yields marginally worse performance for the zero-knowledge (in-domain) attacker, but marginally better performance for the full-knowledge (out-of-domain) attacker. 
Against the pessimistic defender, the pure zero-knowledge attacker performs incredibly well, and using Bayes/Hurwicz somewhat negatively impacts the attacker's interquartile mean reward. 
While the full-knowledge (in-domain) attacker performs worse overall against the pessimistic defender, the pure strategy is better than using Bayes-Hurwicz, as we would expect.

\begin{figure}[h!]
    \centering
    \includegraphics[width=0.85\textwidth]{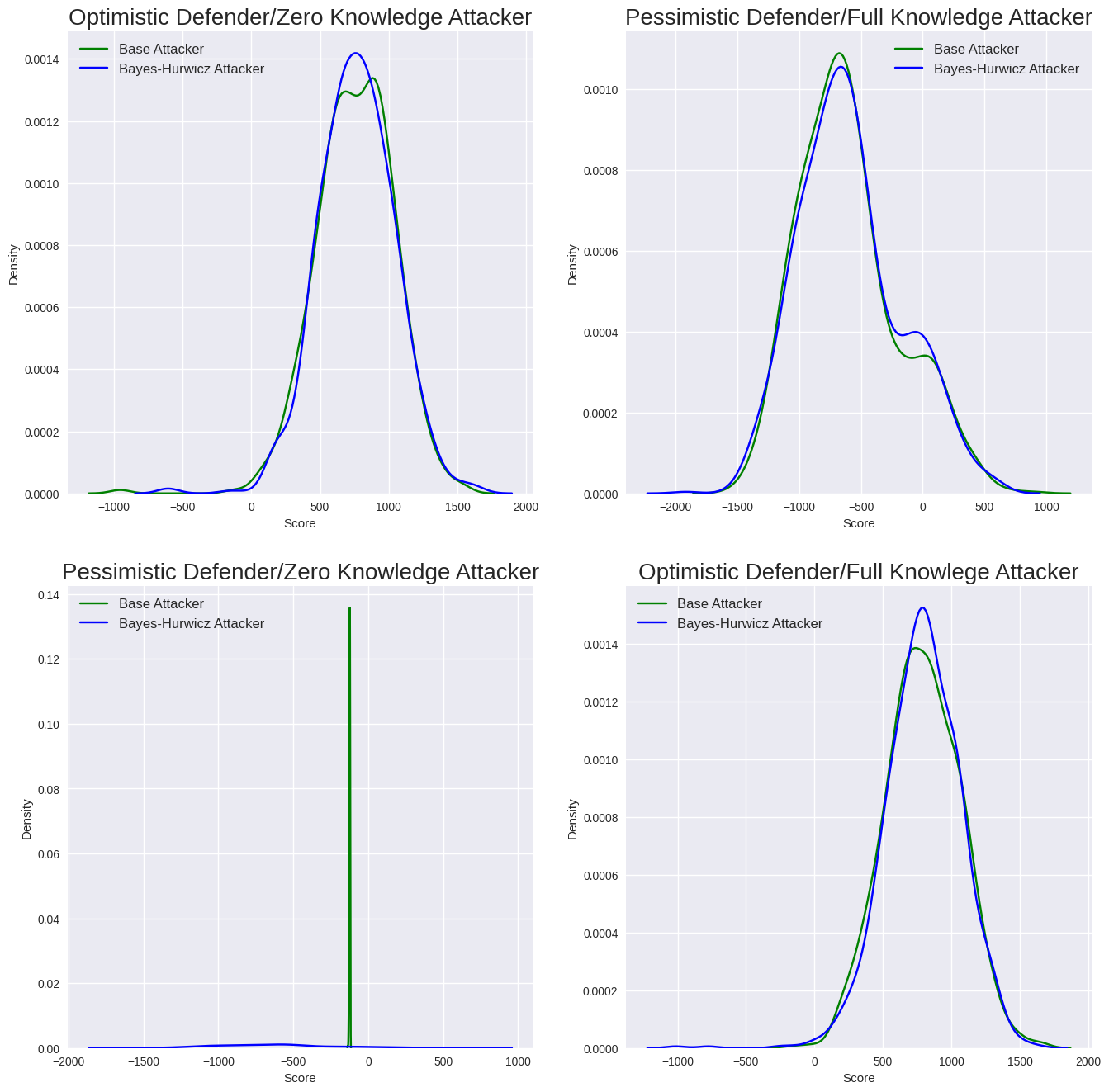}
    \caption{Score distributions for optimistic and pessimistic defenders against base and restricted Bayes/Hurwicz attackers in full-knowledge and zero-knowledge settings. Note that the scale of x and y axes differ among the subfigures.}
    \label{fig:bh_score_dist}
\end{figure}

Results for defenders against both base attackers and those using restricted Bayes/Hurwicz are shown in Figure~\ref{fig:bh_blue}, and reflect the information from Figure~\ref{fig:iqm}.
The optimistic defender experiences a marginal improvement against the restricted Bayes/Hurwicz attackers in both the zero knowledge and full knowledge case.
Meanwhile, the pessimistic defender performs slightly better against the restricted Bayes/Hurwicz attacker in the full knowledge (in-domain) case, but significantly worse against the restricted Bayes/Hurwicz attacker in the zero knowledge (out-of-domain) case. 

\section{Conclusion and Future Work}
The game design assumptions made about an attacker's capabilities and knowledge grow increasingly important as we seek to develop improved mitigation capabilities for cyber defenders.
In this work, we have demonstrated a phenomenon we call ``the price of pessimism'' -- the cost incurred by a defender by assuming a more pessimistic view of what an attacker is likely to know.
Practically, this suggests that the development of defensive agents necessitates careful consideration of what real-world attackers are likely to know, and model those assumptions correctly. 
Specifically, assumptions made about attacker knowledge considerably influence what defenders learn and the efficacy of their response.
In future work, we aim to define more precisely how environmental factors should be determined and the impact of per-node and subnet variability of true and false positive alert rates on both attacker and defender performance.

Our results demonstrate that a defender's assumptions about \textit{a priori} attacker knowledge of an environment have a measurable impact on how that defender responds to potential intrusions.
An assumption that overestimates an attacker's knowledge and the concomitant learned response dynamics from this assumption leads to overreaction to false positives on the part of defending agents, incurring unnecessary costs and leading to poor convergence in reinforcement learning settings.
We conclude that future work in this space seeking to have impact on systems in the real world should account for the likely knowledge and learning dynamics of attackers in addition to those of defenders and should aim to more accurately capture the learning behavior of attackers. 

In future work, we aim to apply these findings to automated threat response by incorporating human factors, threat modeling, and using more complex simulation frameworks.
Although our agents are learning agents, attacks both today and in the foreseeable future are conducted not by pure utility maximizing agents, but by human beings.
As a result, we look to incorporate prospect theory~\cite{tversky1992advances} in future work similar to how such models have been used to align large language models~\cite{ethayarajh2024kto}.
We also aim to explore how incorporating threat intelligence information about sequences of attacker actions and how they lead to different outcomes can constrain the defender's action space, allowing for threat-informed defense.
Finally, given the restricted node states of the YT reinforcement learning environment, research incorporating threat intelligence may need to leverage a simulation framework more similar to real-world environments, like CybORG~\cite{cage_cyborg_2022}.

\bibliographystyle{splncs04}
\bibliography{references}

\appendix

\end{document}